\documentstyle[11pt]{article}
\def\beq{\begin{eqnarray}}    
\def\eeq{\end{eqnarray}}      

\def\al{\alpha}
\def\be{\beta}
\def\ch{\chi}
\def\ga{\gamma}
\def\de{\delta}
\def\vp{\varepsilon}
\def\ep{\epsilon}

\def\ka{\kappa}
\def\la{\lambda}
\def\na{\nabla}

\def\si{\sigma}

\def\ph{\varphi}
\def\ta{\tau}

\def\Ga{\Gamma}

\def\La{\Lambda}

\def\ii{\'{\i}}

\begin{document}
\thispagestyle{empty}

 \vspace*{3mm}
 \begin{center}

 {\LARGE \sl One-loop divergences of quantum gravity }
 \vskip 3mm
 {\LARGE \sl using conformal parametrization}

 \vskip 10mm

 {\bf Guilherme de Berredo-Peixoto}\footnote{ Electronic address:
 peixoto@cbpf.br} \\ Dep. Campos e Part\ii culas (DCP)-Centro Brasileiro
 de Pesquisas F\'{\i}sicas-CBPF
  \\
 \vskip 5mm
 {\bf Andr\'e Penna-Firme}\footnote{ Electronic address:
 andrepf@cbpf.br}
 \\Dep. Campos e Part\ii culas (DCP)-Centro Brasileiro de Pesquisas
 F\'{\i}sicas-CBPF\\
 Faculdade de Educa\c{c}\~ao da Universidade Federal do
 Rio de Janeiro, (UFRJ)\\
 \vskip 5mm
 {\bf Ilya L. Shapiro}\footnote{ Electronic address:
 shapiro@fisica.ufjf.br. On leave from Tomsk Pedagogical
 University, Russia. }
 \\ Departamento de F\'{\i}sica -- ICE, Universidade
 Federal de Juiz de Fora\\
 Juiz de Fora, 36036-330, MG, Brazil

 \end{center}
 \vskip 10mm

\noindent
 {\large \it  Abstract}.$\,\,\,\,$
 {\sl We calculate the one-loop divergences for quantum gravity
 with cosmological constant, using new parametrization of quantum
 metric. The conformal factor of the metric is treated as an
 independent variable. As a result the theory possesses an additional
 degeneracy and one needs an extra conformal gauge fixing. We verify
 the on shell independence of the divergences from the parameter of the
 conformal gauge fixing, and find a special conformal gauge
 in which the divergences coincide with the ones obtained by
 t'Hooft and Veltman (1974). Using conformal invariance of the
 counterterms one can restore the divergences for the conformal
 metric-scalar gravity.}

 \vskip 10mm
 \section{Introduction}

The renormalization of quantum gravity and in particular the 
calculation of
the one-loop divergences for quantum General Relativity is
considered as a problem of special interest. The 
non-renormalizability of quantum gravity has been established 
after the pioneer one-loop calculation
by t'Hooft and Veltman \cite{hove} and Deser and van Nieuwenhuizen
\cite{dene}, who derived the divergences for pure gravity and also 
for the gravity coupled to scalar, vector and spinor fields. 
In both \cite{hove}
and \cite{dene} the background field method has been used such that 
the splitting of the metric was performed according to 
$\,\,\,g_{\mu\nu}\to g_{\mu\nu} + h_{\mu\nu}$. 

Later, the derivations of the one-loop
divergences have been carried out many times, using different
parametrizations of quantum metric and non-minimal gauge fixing 
conditions. The calculations were also done 
for gravity coupled to various kinds of matter fields. One 
can mention: the first calculation for the pure gravity 
in a non-minimal gauge \cite{ktt}; $\,\,\,$
the calculations using plane Feynman diagrams 
with various parametrizations of the quantum metric \cite{capper};
$\,\,\,$
in \cite{local_mom} the result identical to the one of \cite{hove} 
has been achieved using local momentum representation technique; 
$\,\,\,$
the calculation for gravity coupled to Majorana spinor 
using the (slightly modified) Schwinger-DeWitt technique 
\cite{bavi1}; $\,\,\,$ the
calculations in the first order formalism (with affine connection
independent on the metric) using plane
Feynman diagrams \cite{diplom} and background field method and
Schwinger-DeWitt technique \cite{acta}. 
Ref. \cite{acta} contains also the one-loop result
for the $\,g^{\mu\nu}\to g^{\mu\nu} + h^{\mu\nu}\,$ parametrization,
different from the one of \cite{hove}. 
The generalized Schwinger-DeWitt
technique has been applied in \cite{bavi} to confirm the gauge fixing
dependence found in \cite{ktt}. The calculation for gravity with
cosmological constant has been done in \cite{chrisduff} and for the
Einstein-Cartan theory with external spinor current in \cite{EC}.
Recently,
the one-loop calculations for the pure gravity has been performed in
\cite{kalmyk} where the parametrizations like 
$\,g_{\mu\nu}\to g_{\mu\nu} + (-g)^r\,h_{\mu\nu}\,$ 
have been applied. The parametrizations of \cite{kalmyk} are 
more general than the ones used 
in both \cite{hove} and \cite{acta}, so that \cite{kalmyk} 
reproduces both results in the limiting cases.

The interest to the gauge fixing dependence of the
in quantum gravity has been revealed
in the last years, when some more complicated linear gauges 
have been studied \cite{lavrov} (one can consult this 
paper for the list of references concerning the problem
of gauge dependence in quantum field theory and quantum
gravity). The purpose of the present letter is to report
about the calculation of the one-loop divergences
in quantum gravity, in some new parametrization which
is different from those which have been used before.
This parametrization is based on the separation of the
conformal factor from the metric and is related to the
well known conformal structure of gravity (see, for example,
\cite{deser,conf}).  In part, our parametrization resembles 
the one which has been applied in \cite{kani} for the
derivation of the divergences in $\,2+\ep\,$ space-time 
dimensions. As usual, there is the possibility to conduct
an efficient auto-verification of the result, using the on shell
gauge fixing independence. One has to notice that the study 
of conformal gauge in four dimensions has some special importance, 
since its use
permits partial verification of the gauge fixing procedure
$\,h^\mu_\mu=0$, which is usually applied in conformal quantum 
gravity \cite{frts,anmamo}. It is worth to notice that the 
divergences for the Weyl gravity calculated in 
\cite{frts} and \cite{anmamo} differ unlike one uses the
so-called conformal regularization introduced in \cite{frts}. 
The result of our calculation, which is intended
to check the applicability of the conformal gauge 
$\,\,h_\mu^\mu=0$, can be relevant in the general context of 
conformal quantum gravity theories in four dimensions.

The present letter is organized as follows. In the next section
we present the details of the one-loop calculations. The
analysis of the results, including the on-shell
gauge fixing independence is performed in section 3, and in
the last section we draw our conclusions.

\section{One-loop calculation in a conformal gauge}

Our starting point is the gravity action with the
cosmological constant
\beq
S = \frac{1}{\ka^2}\int\; d^4x\sqrt{-g}\, (R+2\La )\,,
\label{iniact}
\eeq
In order to illustrate how the degeneracy related to the
conformal symmetry appears, let us briefly repeat the
consideration of \cite{deser,conf}.

 Performing conformal transformation
 $\,g_{\mu\nu} \to {\hat g}_{\mu\nu}=g_{\mu\nu}\cdot e^{2\si(x)}$,
 one meets relations between geometric
 quantities of the original and transformed metrics:
 \beq
 \sqrt{-{\hat g}} = \sqrt{-g}\;e^{4\si}\,, \;\;\;\;\;\;\;\;\;
 {\hat R} = e^{-2\si}\left[R -
 6\Box\si - 6(\na\si)^2 \right] \,.
 \label{n1}
 \eeq
 Substituting (\ref{n1}) into (\ref{iniact}), after
 integration by parts, we arrive at:
 \beq
 S = \int d^4x \sqrt{-g}\,
 \left\{ \,
\frac{6}{\ka^2}
\,e^{2\si}\,(\na\si)^2
 + \frac{1}{\ka^2}\,e^{2\si}\,R + \frac{2}{\ka^2}\La e^{4\si}\right \},
 \label{n2}
 \eeq
 where $(\na \si)^2
 = g^{\mu\nu}\partial_\mu\si \partial_\nu\si $.
 If one denotes
 \beq
 \ph = \sqrt{{12}/{\ka^2}}\,\cdot\, e^{\si} \,,
 \label{n3}
 \eeq
 the action (\ref{iniact}) becomes
 \beq
 S=\int d^4x \sqrt{-g} \left\{\,
 \frac{1}{2}\,(\na\ph)^2+\frac{1}{12}\,R\,\ph^2+
 \frac{\ka^2}{72}\,\La\,\ph^4\, \right \}\,,
 \label{n5}
 \eeq
that is the action of conformal metric-scalar theory. 
This theory is conformally equivalent to
General Relativity with cosmological constant. Contrary to 
General Relativity, the 
theory (\ref{n5}) possesses extra local conformal symmetry, 
for it is invariant under the transformation
 \beq
 g^{\prime}_{\mu\nu}=g_{\mu\nu}\cdot e^{2\rho(x)}
 \,,\, \, \, \, \, \, \, \, \, \, \,
 \ph^{\prime} = \ph \cdot e^{-\rho(x)}\,.
 \label{nnn}
 \eeq
This symmetry compensates an extra (with respect to
(\ref{iniact})) scalar degree of freedom.

Let us now discuss the relation between two theories on quantum level. In
case of renormalizable field theory the difference between two conformally
equivalent theories appears on quantum level because of conformal anomaly.
For quantum gravity one can not go so far because both theories are
non-renormalizable and therefore anomaly is ambiguous \footnote{For
instance, the divergences of (\ref{iniact}) vanish for the
special gauge fixing, and then anomaly vanishes.}. At the same time,
we can investigate the
difference in quantization of two theories and the resulting
difference in divergences. One has to notice that, despite the 
derivation of
divergences in the theory (\ref{n5}) is possible using the techniques
developed in \cite{bavi} and \cite{odsh}, such a calculation would
be quite difficult. Technically it is much more cumbersome than similar
derivation for the non-minimal, non-conformal metric-scalar theory
\cite{bakaka,spec}. In this paper we do not try to perform this
calculation directly, but instead consider the derivation of the 
one-loop divergences in the theory (\ref{iniact}) using special 
conformal parametrization. 

Since the theory (\ref{iniact}) is diffeomorphism invariant,
it should be quantized as a gauge theory. On the other hand,
the theory (\ref{n5}) has an extra conformal symmetry,
and thus its quantization requires
an extra gauge fixing which is called to remove corresponding
degeneracy. As we shall see later, this is also true for the
quantization of (\ref{iniact}) in conformal variables.

In the framework of the background field method, let us
consider the following shift of the metric
\beq
g_{\mu\nu} \to g^\prime_{\mu\nu} = e^{2\si}\,
[g_{\mu\nu}+h_{\mu\nu}]\,,
\label{param}
\eeq
where $h_{\mu\nu}$ and $\si$ are quantum fields and 
$g_{\mu\nu}$ is the background metric.
All raising and lowering of indices is done
through $g_{\mu\nu}$. The parametrization (\ref{param})
resembles the conformal transformation which led to the
conformal form of the action (\ref{n2}). Then one can expect 
to meet an additional degeneracy for the quantum field, 
related to the conformal symmetry.

For the one-loop divergences, one needs only
the bilinear, in the quantum fields $h^{\mu\nu}$ and $\si$,
part of the action. This part can be presented in the symbolic form:
\beq
S^{(2)}= \int\; d^4x\,\sqrt{-g}\, \left( \, \begin{array}{cc}
h^{\mu\nu} & \si  \end{array}\, \right)\; \hat{H}\; \left( \,
\begin{array}{c} h^{\al\be} \\ \si \end{array}\, \right).
\eeq
Now, one has to introduce the gauge fixing
for the diffeomorphism. We choose the gauge fixing term
in the form
\beq
S_{GF}=-\frac{1}{\al}\int\; d^4x\sqrt{-g}\, \ch _{\mu}\ch ^{\mu}
\label{gafite}
\eeq
with 
\beq
\ch _{\mu}=\nabla _{\al}h^{\al}\mbox{}_{\mu}+
\be\,\nabla _{\mu}h-\ga\,\nabla _{\mu}\si ,
\label{gafipa}
\eeq
where $h=h^{\mu}\mbox{}_{\mu}$ and $\al$, $\be, \ga$ are gauge 
fixing parameters.
It is useful to choose them in such a way that the
bilinear form becomes minimal second order operator.

One can find that this can be achieved by taking
$\al =2$, $\be =-1/2$ and $\ga =2$.  Then the bilinear form
of the action with the gauge fixing term becomes
$$
S^{(2)}+S_{GF}^{(2)}  =  \int\; d^4x\sqrt{-g}\, \left\{
h^{\mu\nu}\,\left[\,
K_{\mu\nu\, ,\,\al\be}(\Box-2\La)+
M_{\mu\nu\, ,\,\al\be}\,\right] \, h^{\al\be}+
\right.
$$
\beq
\left.
+ \si\,\left( \, -4\Box +2R+16\La\,\right) \, \si +  
h^{\mu\nu}\,\left( \,-g_{\mu\nu}\Box
- 2R_{\mu\nu} + g_{\mu\nu}R +
4\La g_{\mu\nu}\, \right) \, \si\, \right\}\,,
\label{minimal_form}
\eeq
where
\beq
K_{\mu\nu\, ,\,\al\be}=\frac{1}{4}\left(\, \de _{\mu\nu\, , \,\al\be}-
\frac{1}{2}g_{\mu\nu}g_{\al\be}\,\right)
\eeq
and
$$
M_{\mu\nu\, ,\,\al\be}= - \frac{1}{4}\,\de _{\mu\nu\, , \,\al\be}R+
\frac{1}{8}\,\left(\,g_{\nu\al}R_{\mu\be}
+g_{\mu\al}R_{\nu\be}+g_{\mu\be}R_{\nu\al}+g_{\nu\be}R_{\mu\al}
\,\right)-
$$
\beq
-\frac{1}{4}\,\left(\,g_{\al\be}R_{\mu\nu}+g_{\mu\nu}R_{\al\be}\,\right)
+\frac{1}{8}\,\left(\,
R_{\mu\al\nu\be}+R_{\nu\al\mu\be}+R_{\nu\be\mu\al}+R_{\mu\be\nu\al}
\,\right)
+\frac{1}{8}g_{\mu\nu}g_{\al\be}R\,,
\eeq
where we have used standard notation
$\,\,\de _{\mu\nu\, , \,\al\be}=\frac{1}{2}(g_{\mu\al}g_{\nu\be}+
g_{\mu\be}g_{\nu\al})$.

It proves useful to
separate the field $h^{\mu\nu}$ into the trace and the traceless
part, $h^{\mu\nu}=\bar{h}^{\mu\nu}+\frac{1}{4}g^{\mu\nu}h$.
Then the bilinear form (\ref{minimal_form}) becomes
\beq
S^{(2)}+S_{GF}^{(2)} & = & \int\; d^4x\sqrt{-g}\, \left\{
\bar{h}^{\mu\nu}\,\left[\, \frac{1}{4}\,\bar{\de}_{\mu\nu\, , \,\al\be}
\,(\Box -2\La )+ M_{\mu\nu\, ,\,\al\be}\,\right] \, \bar{h}^{\al\be}+
\nonumber \right. \\ & + & \left. \bar{h}^{\mu\nu}\,
\left[\,- 2\,R_{\mu\nu}\,\right]\, \si
+ h\, \left[\,-\,\frac{1}{16}\,\Box+
\frac{1}{8}\,\La\,\right]\, h + \nonumber \right. \\ & + & \left.
h \,\left[\,-\Box + \frac{1}{2}\,R+4\La \,\right]\, \si +
\si\, \left(\, -4\Box + 2R + 16\La \,\right]\, \si
\, \right\}\,.
\label{degen}
\eeq
Here 
$$
{\bar \de}_{\mu\nu , \al\be} = \de_{\mu\nu , \al\be} 
- \frac14\,g_{\mu\nu} g_{\al\be}
$$
is the projector to the traceless states. 
The expression (\ref{degen}) exhibits the degeneracy in the mixed 
$\,\,h-\si\,\,$
sector, and hence further calculation requires some
additional restriction on the quantum fields. This degeneracy
is a direct consequence of the conformal symmetry (\ref{nnn})
and thus we have to fix this symmetry. Let us choose the
conformal gauge fixing in the form $\si =\la h$ with $\,\la\,$
being the gauge fixing parameter. Then (\ref{degen}) becomes:
\beq
S^{(2)}+S_{GF}^{(2)} & = & \int\; d^4x\sqrt{-g}\, \left\{\,\,
\bar{h}^{\mu\nu}\,\left[\, \frac14\,\bar{\de}_{\mu\nu\, , \,\al\be}
\,(\Box -2\La)+ M_{\mu\nu\, ,\,\al\be}\,\right] \,\bar{h}^{\al\be}+
\nonumber \right. \\ & + & \left. \bar{h}^{\mu\nu}\,
\left[\,-2\la R_{\mu\nu}\,\right]\, h + h\,\left[\,
\, b_{1}\Box +2b_{2}\La + b_{3}R \,\right)\, h\,\,\right\}
\label{hgrav}
\eeq
where we introduced the notations
\beq
b_1 = -\frac{1}{16}-\la - 4\la ^2; \;\;\;\; \;\;\;\;
b_2=\frac{1}{16}+2\la +8\la ^2; \;\;\;\;\;\;\;\;
b_3=\frac{1}{2}\la +2\la ^2.
\eeq

The total one-loop divergences will be given by
\beq
\Ga ^{(1)}_{{\rm div}}=\frac{i}{2}
Tr\,{\rm ln}\, \hat{H}_{{\rm grav}}|_{{\rm div}}
-iTr\,{\rm ln}\, \hat{M}|_{{\rm div}}
\eeq
where the last term is the contribution
from the ghost fields, and $\hat{H}_{{\rm grav}}$ is the operator
corresponding to eq. (\ref{hgrav}). The standard
Schwinger-DeWitt algorithm enables one to derive
$$
\frac{i}{2}
Tr\,{\rm ln}\, \hat{H}_{{\rm grav}}|_{{\rm div}}
= -\frac{1}{\vp}\int\; d^4x\sqrt{-g}\, \left\{\,
\frac{19}{18}R^2_{\rho\la\si\ta}+
\left( \frac{4}{b_1}\la ^2-\frac{55}{18}\right)\,
R^2_{\rho\la}+
\right.
$$
\beq
\left.
\left( \frac{59}{36}
-\frac{\la ^2}{b_1}+\frac{b_3}{6b_1}+\frac{b_3^2}{2b_1^2}
\right)\, R^2+
\left(\frac{2b_2b_3}{b_1^2}+\frac{b_2}{3b_1}+9\right)\, R\La+
\left(\frac{2b_2^2}{b_1^2}+18\right)\, \La ^2 \right\}
\label{newforma}
\eeq
where $\,\vp=(4\pi )^2(n-4)$.
Also, the operator of the ghost action $\hat{M}$ is
\beq
\hat{M}_\mu^\nu \,=\, - \,\de_\mu^\nu\,\Box \,-\,R_\mu^\nu \,.
\label{ghost_oper}
\eeq
We remark that the ghost operator does not depend
on the gauge transformation of the field $\si$, because
at the one-loop level, in the background field method,
the generator of the gauge transformations
which enters into the expression for $\,\hat{M}_\mu^\nu \,$ 
is the one for the background (not quantum!) fields
\cite{hove} (see also \cite{book}) and in case of
$\si$ this operator is zero.

Calculation of the ghost contribution yields standard result \cite{hove}
\beq 
-iTr\,{\rm ln}\, \hat{M}|_{{\rm div}} & = & \frac{1}{\vp}\int\;
d^4x\sqrt{-g}\, \left\{\, -\frac{11}{90}\, E +\frac{7}{15}\,
R^2_{\mu\nu}+\frac{17}{30}\, R^2\,\right\} , 
\eeq 
where
$E=R^2_{\mu\nu\al\be}-4R_{\mu\nu}^2+R^2$. 
Finally, one arrives at the
following one-loop divergences: 
\beq 
\Ga ^{(1)}_{{\rm div}}
=-\frac{1}{\vp}\int\; d^4x\sqrt{-g}\, \left\{\,
p_1(\la)E+p_2(\la)C^2+p_3(\la)R^2+p_4(\la)R\La +p_5(\la)\La ^2 \right\}
\label{naschitali} 
\eeq 
where $C^2$ is the square of the Weyl tensor
$\,C^2=E+2(R_{\mu\nu}^2-\frac13\,R^2)\,$ and \beq p_1(\la) & = &
\frac{1}{180}\, \frac{149+2384\la +15296\la ^2} {(1+8\la )^2}, 
\nonumber
\\
p_2(\la) & = & \frac{1}{20}\, \frac{7+112\la -192\la ^2} {(1+8\la )^2},
\nonumber \\ p_3(\la) & = & \frac{1}{12}\, \frac{3+80\la +1152\la
^2+6144\la ^3 +10240\la ^4} {(1+8\la )^4}, \nonumber \\ p_4(\la) & = &
\frac{2}{3}\, \frac{13+432\la +5696\la ^2 +31744\la ^3+63488\la ^4}
{(1+8\la )^4}\;\;\; {\rm and} \nonumber \\ p_5(\la) & = & 4\,
\frac{5+176\la +2368\la ^2 +13312\la ^3+26624\la ^4} {(1+8\la )^4}.
\label{1-loop_divs} 
\eeq 
The above expression (\ref{naschitali}),
(\ref{1-loop_divs}) contains complicated dependence on the gauge fixing
parameter $\,\la\,$. Besides, the one-loop divergences may depend on
others
gauge fixing parameters $\,\al,\,\be,\,\ga\,$ from (\ref{gafipa}). Here we
are interested only in the dependence on $\,\la$, and keep
$\,\al,\,\be,\,\ga\,$ fixed as before.

\section{Analysis of the results}

The expression (\ref{naschitali}), (\ref{1-loop_divs}) looks quite
cumbersome and somehow chaotic because of the complicated dependence on
the
gauge fixing parameter $\la$. But, in fact, there are a few possibilities
to check and analyze it. First of all, for the value $\la=0$, all the
$\si$-field contributions drop and we arrive at the well-known result
\cite{hove,chrisduff} 
\beq 
\Ga ^{(1)}_{{\rm div}} =-\frac{2}{\vp}\int\;
d^4x\sqrt{-g}\, \left\{\,
\frac{1}{120}R^2+\frac{7}{20}R_{\mu\nu}^2+\frac{53}{90}E +
\frac{13}{3}\,R\La + 10\,\La^2 \,\right\}\,. 
\label{hv} 
\eeq 
For other
values of $\la$ the divergences are different and one can check that the
$\la$-dependence can not be compensated by the change of other gauge
fixing
parameters $\,\al,\be\,$ or by the change of parameter $\,r\,$ introduced
in \cite{kalmyk}.

If we take a limit $\la\to\infty$, the
result is not conformal invariant, as one could naively
expect. Let us give some additional comment on this point.
The above calculation
can be regarded as a particular case of the much more complicated
derivation of the one-loop divergences in the theory (\ref{n5}),
which was mentioned in the Introduction.
In general calculation one is supposed to shift both fields
$\ph$ (or $\si$, this is equivalent)
and $g_{\mu\nu}$, while in this paper we took the
background scalar to be constant. Let us imagine, for a moment, that 
we shifted both fields
\beq
\si \to \chi + \si\,,\,\,\,\,\,\,\,\,\,\,\,
g_{\mu\nu} \to g_{\mu\nu} + h_{\mu\nu}\,.
\label{total_shift}
\eeq
As far as we believe into conformal invariance of the one-loop
divergences\footnote{Some remarkable example of the opposite one
can meet in the Weyl gravity, where the results of two 
one-loop calculations \cite{frts} and \cite{anmamo} coincide 
only after the use of the so-called conformal regularization 
\cite{frts}. The lack of equivalence between the results
of \cite{bakaka} and \cite{spec} may indicate to the 
similar problem.},
the result for conformal metric-scalar theory is \cite{alter}
$$
\Ga_{div}=\frac{1}{\ep}\,\int d^{4}x\sqrt{-g}\left\{ p_1\,E
+ p_2\,C^{2} +
\right.
$$
\beq
\left.
+ p_3\left[ R - \frac{3{\Box }\zeta}\zeta
+\frac{3({\nabla }\zeta )^2}{2\zeta ^2}
\right]^2
+ p_4\zeta\,\left[ R - \frac{3{\Box }\zeta}{\zeta }
+\frac{3(\nabla \zeta )^{2}}{2\zeta ^{2}}\right]
+ p_5\zeta^2
\,\right\},
\label{conf2}
\eeq
where $\zeta =\zeta (\chi )$ is some function of $\,\chi$.
The procedure accepted in this paper is equivalent to taking 
$\,\chi=const$, and therefore
(\ref{1-loop_divs}) should be regarded as (\ref{conf2}) with
constant $\zeta$. Obviously, constant $\zeta$ does not transform
and the conformal invariance is lost.

In order to verify the result of the calculation, one can use classical
equations of motion $R_{\mu\nu} = -\La\,g_{\mu\nu}$.
On shell the divergences become 
\beq 
\Ga^{(1)\,{\rm
on-shell}}_{{\rm div}}=-\frac{1}{\vp}\int\; d^4x\sqrt{-g}\, \left\{\,
\frac{53}{45}E-\frac{58}{5}\La ^2 \right\}\,, 
\label{onshell} 
\eeq
independent on the gauge fixing parameter $\la$. As a consequence, 
the on shell renormalization group equation for the dimensionless 
cosmological constant $\,\ka^2\Lambda\,\,$
\cite{frts} is gauge fixing independent in our conformal
parametrization. One has to notice that the coefficients of 
(\ref{onshell})
are linear combinations of all five functions (\ref{1-loop_divs}), and
thus the complete cancellation of the $\la$-dependence, together with 
(\ref{hv}), provide a very confident verification of the result
(\ref{naschitali}).

\section{Conclusions}

We have studied the equivalence between General Relativity and conformal
metric-scalar theory on quantum level. The one-loop divergences were
calculated for quantum gravity, for the first time this was done in the
conformal parametrization for quantum metric. We have found that the
dependence on the new gauge fixing parameter disappears on shell. This
gives an efficient check to the whole procedure based on fixing the
conformal symmetry by using the trace of quantum metric $h=h_{\mu}^{\mu}$.
The results of our work show that the source of the discrepancy 
in the results for the quantum Weyl gravity is not caused by this
conformal gauge fixing. Finally, 
the supposition of conformal invariance of the counterterms enables one to
restore the divergences for the gravity coupled to conformal scalar field
(\ref{conf2}).

\vskip 5mm \noindent {\bf Acknowledgements}. I.Sh. and G.B.P. are grateful
to CNPq for for permanent support and for the scholarship correspondingly.
A.P.F. is grateful to DCP-CBPF for hospitality and support.

\vskip 10mm

\end{document}